\documentclass[%
 reprint,
 amsmath,amssymb,
 pra,
]{revtex4-2}

\usepackage{graphicx}
\usepackage{dcolumn}
\usepackage{bm}
\usepackage{color}

\begin{document}

\preprint{PRA/123-QED}

\title{Interaction-enhanced transmission imaging with Rydberg atoms}

\author{Xiaoguang Huo$^{1}$, J. F. Chen$^{2,3}$, Jing Qian$^{1,
*}$, Weiping Zhang$^{4,5,6}$}%
\affiliation{$^{1}$State Key Laboratory of Precision Spectroscopy, Department of Physics, School of Physics and Electronic Science, East China
Normal University, Shanghai 200062, China}
\affiliation{$^{2}$Guangdong Provincial Key Laboratory of Quantum Science and Engineering, Southern University of Science and Technology, Shenzhen, 518055, China}
\affiliation{$^{3}$Shenzhen Institute for Quantum Science and Engineering, Southern University of Science and Technology, Shenzhen, 518055, China}
\affiliation{$^{4}$School of Physics and Astronomy, and Tsung-Dao Lee Institute, Shanghai Jiao Tong University, Shanghai, 200240, China}
\affiliation{$^{5}$Shanghai Research Center for Quantum Science, Shanghai, 201315, China}
\affiliation{$^{6}$Collaborative Innovation Center of Extreme Optics, Shanxi Universty, Taiyuan, Shanxi 030006, China}

\email{jqian1982@gmail.com}

\date{\today}

\begin{abstract}
Atomic-scale imaging offers a reliable tool to directly measure the movement of microscopic particles. We present a scheme for achieving a nondestructive and ultrasensitive imaging of Rydberg atoms within an ensemble of cold probe atoms. This is made possible by the interaction-enhanced electromagnetically induced transparency at off-resonance which enables an extremely narrow absorption dip for an enhanced transmission. Through the transmission of a probe beam, we obtain the distribution of Rydberg atoms with both high spatial resolution and fast response, which ensures a more precise real-time imaging. Increased resolution compared to the prior interaction-enhanced imaging technique allows us to accurately locate the atoms
by adjusting the probe detuning only. This new type of interaction-enhanced transmission imaging can be utilized to other impure systems containing strong many-body interactions, and is promising to develop superresolution microscopy of cold atoms.
\end{abstract}

\maketitle


\section{INTRODUCTION}

The demand for imaging individual Rydberg atoms with high spatial and temporal resolutions gave birth to the development of versatile optical imaging techniques. Earlier methods accessible for that purpose were based on, {\it e.g.}, the field ionization imaging \cite{PhysRevLett.107.103001,PhysRevA.87.013407} or fluorescence imaging \cite{2012Observation,2016Tunable} of atoms uncovering both virtues. 
However these detection methods are destructive and the atoms cannot be re-used. For showing important applications in diverse areas, such as the quantum information processing \cite{RevModPhys.82.2313} and the precision measurement \cite{giovannetti_lloyd_maccone_2011}, a nondestructive and high-efficiency detection of Rydberg states is imperative. Recently a superconducting microwave cavity has been used for efficient single-shot nondestructive measurement of Rydberg-atom ensembles enabled by its enhanced sensitivity \cite{PhysRevLett.123.193201}, and a pulsed ion microscope was just reported for an achievable resolution below 200 nm \cite{PhysRevX.11.011036}. These achievements
open up new perspectives for imaging Rydberg atoms.

Alternatively a promising approach proposed by refs. \cite{PhysRevA.84.041607,PhysRevLett.108.013002}, is the interaction enhanced imaging (IEI) which manifests as a nondestructive and state-selective optical detection of strongly interacting impurities. Over the last decade IEI has been actively pursued by experimental devotion in versatile systems \cite{Guenter2013Observing,2016Interaction} as a new protocol for investigating the readout of the time-resolved dynamics of ions \cite{PhysRevLett.124.053401}, molecules \cite{Huo_2017} or Rydberg qubits \cite{PhysRevLett.127.050501}, the generation of single-photon transistor with high gain \cite{Gorniaczyk2016Enhancement} and various long-range interactions \cite{PhysRevLett.121.193401,2019Proposal}. IEI underlyingly relies on the strong impurity-probe interaction in the vicinity of each impurity, which can induce an enhanced absorption imaging in the case of resonant electromagnetically induced transparency (EIT) \cite{2012Electromagnetically,Yi2020Experimental}. EIT spectra enable a direct nondestructive detection of highly-excited Rydberg levels 
\cite{PhysRevLett.98.113003}. And the interaction induced Rydberg blockade has found wide applications in quantum information processing \cite{PhysRevLett.112.073901}.
By mapping this strong Rydberg-Rydberg interaction onto the light field one can resolve the property of each impurity without destroying it. However as for Rydberg impurities, their spatial response range is limited by the Rydberg blockade radius, typically around a few micrometers \cite{PhysRevLett.105.193603,PhysRevLett.107.213601}. Reducing it down to the level of sub-micrometer-scale has to involve a strong coupling laser, which is still challenging for current experimental implementation.

\begin{figure*}
\centering
\includegraphics[width=4.1in,height=2.1in]{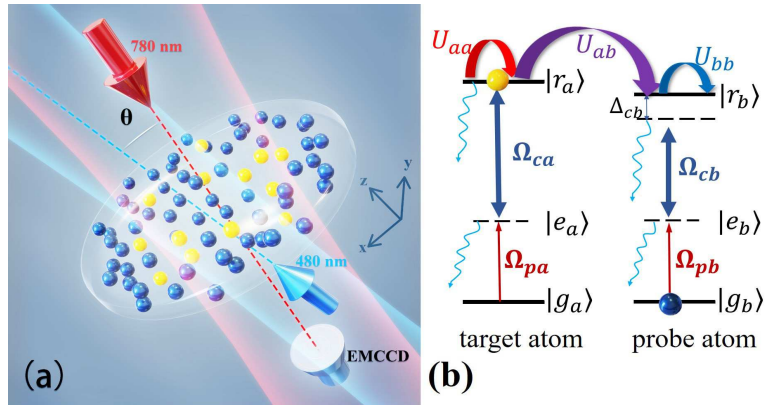}
\caption{(color online)(a) Sketch for the interaction-enhanced transmission imaging. Target atoms(yellow) are randomly embedded within an ensemble of dense quasi-two-dimensional probe atoms(blue) in $x$-$y$ plane. The probe atoms interact with two light fields (probe and coupling) propagating in $ -z$ and $+z$ directions respectively, which are illuminated via a two-photon transition from the ground state $|g_{b}\rangle=|5S_{1/2}\rangle$ to the Rydberg state $|r_b\rangle=|80S_{1/2}\rangle$. The preparation of the Rydberg state $|r_a\rangle=|85S_{1/2}\rangle$ of target atoms is also enabled by a two-photon excitation.
The probe and coupling lasers with wavelengths 780 nm and 480 nm, can induce a negligible Doppler shift as they are almost vertical to the stochastic atomic movements in $x$-$y$ plane.
By detecting the probe-field transmission at the end of $-z$ axis with a cooled EMCCD camera, the position of all target atoms could be precisely resolved from the formation of a narrow transmission ring. In order to reduce the scattering of the coupling laser, a tiny separation angle $\theta\leq5^\textbf{o}$ between the probe and coupling beams is designed. (b) Relevant energy levels and atom-field interactions. The strong probe-target interactions $U_{ab}$ will overcome the big detuning $\Delta_{cb}$ with respect to state $|r_b\rangle$, arising an enhanced  transmission rate at a specific probe-target spacing. $U_{aa}$ and $U_{bb}$ reflect the strength of target-target interactions and probe-probe interactions.}
\label{fig:model}
\end{figure*}

In the present work we develop the protocol of IEI by placing it in an off-resonant EIT environment \cite{PhysRevLett.114.123005}, which benefits from a very narrow absorption dip instead of a broad absorption peak, consequently named as {\it interaction-enhanced  transmission imaging} (IETI). This dip is caused by the compensation between a big probe-atom detuning at off-resonance and the strong impurity(target)-probe interaction, whose width can be flexibly adjusted to be orders of magnitude narrower than a typical blockade radius of a few $\mu$m. By carrying out such an off-resonant measurement with respect to the probe atoms, it is feasible to obtain an ultraprecise imaging of random Rydberg targets with its spatial resolution increased by one order of magnitude as compared to the IEI technique, which is also accompanied by a fast optical response in the real-time detection. Increased resolution in IETI can facilitate an accurate and quick positioning of atoms, especially when multi-Rydberg atoms overlap within a specific region. The scheme is deserving a future experimental exploration for the realization of superresolution microscopy of cold atomic ensembles.

\section{Theoretical strategy}
Principle of our approach illustrated in Fig.\ref{fig:model} presents that, a few target atoms surrounded by a large number of background probe atoms, 
are randomly embedded in a quasi-two-dimensional system. Initially, given the ground target atoms, the probe atoms coupled by light fields $\Omega_{pb}$ and $\Omega_{cb}$ will suffer from an inefficient excitation due to the presence of a large off-resonant detuning $\Delta_{cb}\gg\Omega_{pb},\Omega_{cb}$ with respect to $|r_b\rangle$.
However, once the target atoms are excited to the Rydberg state $|r_a\rangle$ through a two-photon optical pumping with Rabi frequencies $\Omega_{pa}$ and $\Omega_{ca}$, the induced probe-target interaction $U_{ab}$ acts as an effective detuning for the detuned level $|r_b\rangle$. Here $U_{ab}=C_6^{ab}/R^6$ forms as a van der waals({\it vdWs})-type for the {\it S-S} state interaction between two Rydberg levels, where $C_6^{ab}$ is the interaction coefficient and $R$ reflects the probe-target distance \cite{PhysRevLett.105.193603}. The presence of $U_{ab}$ possibly overcomes $\Delta_{cb}$ with respect to the probe atom, and hence gives rise to a spatial EIT effect \cite{gavryusev}. Therefore the position of such randomly-distributed target atoms can be spatially resolved by utilizing a very narrow EIT transmission window at off-resonance \cite{PhysRevLett.114.203002}.

To describe the probe-atom absorption that undergoes an off-resonant EIT excitation, the Hamiltonian describing a single {\it probe} atom is given by
\begin{equation}
\mathcal{H}_b=-\frac{1}{2}(\Omega_{pb}|e\rangle\langle g|_b+\Omega_{cb}|r\rangle\langle e|_b+H.c.) 
+\Delta^{\prime}_{cb}|r\rangle\langle r|_b.
\end{equation}
The presence of an excited target atom will cause a finite energy shift $U_{ab}$ to state $|r_b\rangle$, which has been translated into an effective two-photon detuning of the probe atoms $\Delta^{\prime}_{cb}=\Delta_{cb}+U_{ab}$. For probe atoms, a typical excitation probability is very small due to the off-resonant detuning $\Delta_{cb}$ as well as $\Omega_{pb}\ll\Omega_{cb}$, making the probe-probe interaction $U_{bb}$ negligible. However in the resonant case where $\Delta_{cb}$ has been overcome by $U_{ab}$ at a certain distance, $U_{bb}$ adding to the two-photon detuning $\Delta_{cb}^{\prime}$ may become an important factor during measurement [Appendix B].
On the other hand, the target-target interaction $U_{aa}$ plays a role only if more than two target atoms are coupled to state $|r_a\rangle$ at the same time, and we will consider this effect in the study of many-atom imaging.

Here we begin with a detailed analysis for the case of single target atom. The evolution of the density matrix $\rho_b$ of probe atom is governed by the master equation,
\begin{eqnarray}
\dot{\rho}_b=-i[\mathcal{H}_b,\rho_b]+\mathcal{L}_b, \label{ms}
\end{eqnarray}
with the Liouville operator $\mathcal{L}_b$ expressed as \cite{Raitzsch_2009},
\begin{eqnarray}
\mathcal{L}_b=
\begin{pmatrix}
\Gamma_{eg}\rho_{b,ee} & -\frac{1}{2}\gamma_{2}\rho_{b,ge} & -\frac{1}{2}\gamma_{3}\rho_{b,gr} \\
-\frac{1}{2}\gamma_{2}\rho_{b,eg} & -\Gamma_{eg}\rho_{b,ee}+\Gamma_{re}\rho_{b,rr} &
-\frac{1}{2}(\gamma_{2}+\gamma_{3})\rho_{b,er} \\
-\frac{1}{2}\gamma_{3}\rho_{b,rg} & -\frac{1}{2}(\gamma_{2}+\gamma_{3})\rho_{b,re} & -\Gamma_{re}\rho_{b,rr}
\end{pmatrix}\nonumber
\end{eqnarray}
where $\rho_{b,ij}$ stands for the matrix element and the subscript $b$ represents the probe atom. $\Gamma_{ij}$ is the rate of population decay from state $|i\rangle$ to state $|j\rangle$ $(i,j=g,e,r)$. $\gamma_2=\gamma_e+\Gamma_{eg}$ and $\gamma_3=\gamma_r+\Gamma_{re}$, where $\gamma_e$ and $\gamma_r$ refer to the dephasing rate due to the loss of coherence in the atomic elastic collisions and other dynamics, which are not associated with the population transfer \cite{boyd2020nonlinear}.

\begin{figure}
\includegraphics[width=0.49\textwidth]{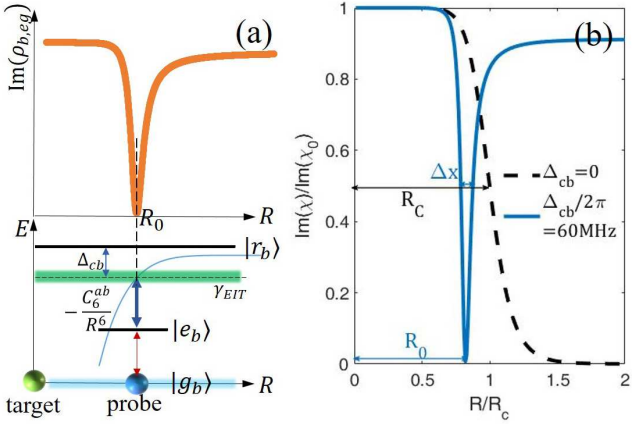}
\caption{(a) Imaginary part of $\rho_{b,eg}(r)$ represents the probe absorption. At a certain distance with $\delta(r)=0$, it arises a critical distance $R_0 \approx \sqrt[6]{C_6^{ab}/\Delta_{cb}}$ at which a narrow absorption dip emerges. (b) The absolute absorption $\text{Im}[\chi]$ normalized by $Im(\chi_0)=2n_0\mu^2_{eg}/(\epsilon_0\hbar\gamma_{2})$ {\it vs} the relative spacing $R$(in unit of $R_c=\sqrt[6]{2\gamma_{2}C_6^{ab}/\Omega^2_{cb}}$) under various $\Delta_{cb}$ values. For $\Delta_{cb}=0$ there exists a wider absorption-enhanced peak(black-dashed) within the Rydberg-blockade volume $R_c$, as similar as Fig.2(b) of ref.\cite{PhysRevLett.108.013002}. However if $\Delta_{cb}\neq0$ a narrower absorption dip(blue-solid) emerges, which ensures a strong transmission signal due to $\gamma_3\ll\Omega_{cb}$. In plotting (b) we set   
$\Omega_{cb}/2\pi=15$ MHz,  $\Omega_{pb}/2\pi=0.15$ MHz, $\gamma_2/2\pi=6.1$ MHz, $\gamma_3/2\pi=10$ kHz, $C_6^{ab}/2\pi=204.8$ GHz$\cdot\mu m^6$.} 
\label{fig:absorption}
\end{figure}

By taking $\dot{\rho}_b=0$ in Eq.(\ref{ms}) we obtain the stationary element $\rho_{b,eg}$ for describing the probe transition \cite{PhysRevLett.74.666},
\begin{eqnarray}
\rho_{b,eg}(r,v)dv=\frac{\Omega_{pb}(2\delta(r)-i\gamma_{3})}{\Omega^2_{cb}+i\gamma_{2}(2\delta(r)-i\gamma_{3})}\mathcal{N}(v)dv,
\label{rhoeg}
\end{eqnarray}
under the assumptions of $\Omega_{pb}\ll\Omega_{cb}$, $\rho_{b,gg}\approx 1$, $\rho_{b,ee}\approx 0$ and $\rho_{b,rr}\approx 0$. Remarkably in Eq.(\ref{rhoeg}) the new denotation $\delta(r)$ is expressed as
\begin{equation}
    \delta(r)=\Delta_{cb}-\frac{C_6^{ab}}{R^6}+U_{bb} - (\vec{k}_c+\vec{k}_p)\cdot \vec{v},
\label{drr}
\end{equation}
where the probe-target spacing $R=|r-r_a|$, represents the effective two-photon detuning and $r_a$ is the random location of a  stationary target atom. For a moving target atom $r_a$ is also time dependent [see Eq.(\ref{ri})].
Accounting for the thermal motion of probe atoms that translates into a Doppler frequency shift to the atomic internal levels we phenomenologically introduce this effect where $\mathcal{N}(v)$ in Eq.(\ref{rhoeg}) is the number density of $^{87}Rb$ probe atoms with velocity $\vec{v}_b$ and $\vec{k}_{c,p}$ is the wavevectors of the coupling or probe beam. $\mathcal{N}(v)$ takes the form of a Maxwell-Boltzmann function at thermal equilibrium. To reduce the impact of Doppler shift we design the two beams propagate nearly vertical to the atom movement, as shown in Fig.\ref{fig:model} (a).
To this end, the susceptibility $\chi(r,v)$ corresponding to the atomic transition driven by the probe field is given by
\begin{equation}
\chi(r,v)dv=\frac{2\mu_{ge}}{\epsilon_0E_p}n(r)\rho_{b,eg}(r,v)dv,
\label{chir}
\end{equation}
where the Gaussian atomic number density is $n(r) = n_0\exp(-r^2/2\sigma_r^2)$ with $n_0$ the peak and $\sigma_r$ the half-width. $E_p$ is the probe electric field amplitude, $\mu_{ge}$ is the transition dipole moment, and $\epsilon_0$ is vacuum dielectric constant. The real (imaginary) part of the susceptibility $\chi(r,v)$ corresponds to the dispersion (absorption) of the probe light, caused by the atomic medium. By integrating the imaginary part of Eq.(\ref{chir}) over the velocity distribution for atomic temperature $T_0$ we can obtain an absolute absorption coefficient Im$(\chi(r))$ as a function of $R$.

Figure \ref{fig:absorption} illustrates the probe absorption by calculating the imaginary part. First we understand the essence of transmission-enhanced imaging by following the map of Fig.\ref{fig:absorption}(a). In the vicinity of an off-resonant driving probe atom, detuned by $\Delta_{cb}$ with respect to a Rydberg state $|r_b\rangle$, the energy level of $|r_b\rangle$ is also shifted by the strong probe-target potential \textcolor{black}{$U_{ab}=C_6^{ab}/R^6$}. Here $r_a=0$ and $R=|r|$ are assumed. If satisfying \textcolor{black}{ $\Delta_{cb}-C_6^{ab}/R^6\approx0$} and $\Delta_{cb}\neq0$ one can obtain a narrow absorption dip at \textcolor{black}{$R_0 \approx \sqrt[6]{C_6^{ab}/\Delta_{cb}}$} with its width $\Delta x$ much smaller than the blockade radius $R_c$. 
The strength of the Doppler shift $(\vec{k}_c+\vec{k}_p)\cdot \vec{v}_b$ can be made orders of magnitude smaller than the Rydberg shift $U_{ab}$. For example a rough estimation based on the wavevector $|\vec{k}_c|=13.09$ $\mu m^{-1}$($\lambda_c=480$ nm), $|\vec{k}_p|=8.06$ $\mu m^{-1}$($\lambda_p=780$ nm) and the most probable speed $v_{mps}=\sqrt{2kT_0/m}\approx 4.37$ $cm/s$ at $T_0=10$ $\mu$K, gives rise to a maximal value for describing the Doppler shift, which is $(|\vec{k}_c|+|\vec{k}_p|)v_{mps}\approx 0.93$ MHz $\ll |U_{ab}|$. In fact
$\vec{k}_{c}$ and $\vec{k}_{p}$ are counter-propagating and almost vertical to the atomic velocity $\vec{v}_b$ in our scheme, leading to a perfect Doppler-free measurement.
So \textcolor{black}{$\delta(r)\approx \Delta_{cb}-C_6^{ab}/R^6$} is confirmed. Note that we also ignore the probe-probe interaction $U_{bb}$ due to the poor probe excitation probability within the Rydberg EIT region for $\Omega_{pb}\ll\Omega_{cb}$. The influence of $U_{bb}$, especially at the position of absorption dip has been explicitly discussed in Appendix B. The resulting enhanced probe transmission can precisely reflect the position of the target atom. This high-contrast and high-precise transmission signal could be suited for the target-atom imaging. 

However if $\Delta_{cb}=0$ as in traditional IEI schemes, the absorption response manifests as an opposite change. As $R\to \infty$ the excited state $|r_{b}\rangle$ becomes resonant and suffers from a zero absorption. Yet within the blockade radius $R_c$, {\it i.e.}, \textcolor{black}{$R<R_c = \sqrt[6]{2\gamma_{2}C_6^{ab}/\Omega^2_{cb}}$}, an enhanced absorption can return a well signature for imaging the location of the target Rydberg atom, although the spatial resolution is relatively poor. Because the broad absorption which is restricted by $R_c$ will make the position measurement insensitive. As a consequence it is insufficient for achieving ultraprecise microscopic imaging of individual atoms by using traditional IEI technology.

A quantitative verification for different probe absorption rates is comparably illustrated in Fig.\ref{fig:absorption}(b).
It is clearly shown that the resonant case of $\Delta_{cb}=0$(black-dotted) allows for an absorption-enhanced signal with its broad half-width $R_c$ at half-maximum. However our approach using $\Delta_{cb}/2\pi = 60$ MHz(\textcolor{black}{blue-solid}) greatly benefits from a higher spatial resolution, characterized by an extremely narrow width $\Delta x$, which is given by\textcolor{black}{
\begin{equation}
\Delta x\approx(\frac{C_6^{ab}}{\Delta_{cb}-\alpha})^{1/6}-(\frac{C_6^{ab}}{\Delta_{cb}+\alpha})^{1/6}
\label{deltax}
\end{equation}}
occurring at the absorption dip $R=R_0$ \textcolor{black}{with $\alpha=\sqrt{\Omega_{cb}^4-\gamma_2^2\gamma_3^2}/(2\gamma_2)$}. At that place an approximately 100$\%$ transmission probability can be obtained. $\Delta x$ is easily tunable by the detuning $\Delta_{cb}$, and in principle \textcolor{black}{ $\Delta_{cb}\gg\alpha$} will lead to arbitrary scale of spatial resolution because $\Delta x\to 0$ as long as the contrast of images permits. Such a narrower absorption dip could deeply improve the imaging precision for the target atoms, promising for the development of superresolution technology.

\section{Single-target-atom imaging}

\subsection{High spatial resolution}

\begin{figure}
\centering
\includegraphics[width=3.5in,height=2.9in]{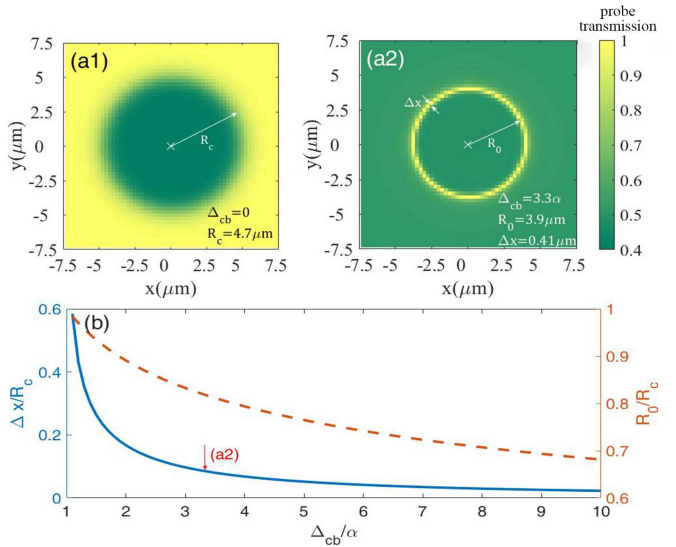}
\caption{Transmission images of single target atom surrounded by a dense ensemble of probe atoms. The unique target atom is located at the center $x=y=0$, denoted by a white cross. In (a1), by using $\Delta_{cb}=0$ as in earlier IEI schemes, the spatial resolution $R_c$ decided by the half-width of transmission, is smoothly increasing as $r=\sqrt{x^2+y^2}$ grows; in contrast, (a2) shows a bright and narrow transmission ring at $r=R_0$ with $\Delta_{cb}=3.3\alpha$, which serves as a more precise determination for the position of target atom, because of its higher spatial resolution $\Delta x \approx R_c/10$. (b) The ratio $\Delta x/R_c$(blue-solid) {\it vs} $\Delta_{cb}$ represents a significant decreasing tendency. \textcolor{black}{The red-dashed line shows that the ratio of $R_0/R_c$ continuously decreases and is always smaller than 1.0 as $\Delta_{cb}$ increases. Red arrow points to the parameters used in (a2).}}
\label{fig:singleatom}
\end{figure}

To carry out numerical calculations for the EIT imaging of random-embedded target atoms, all the probe-target, probe-probe and target-target interactions have to be considered. This is a many-body problem. In our calculation, to prepare multi-target atoms on Rydberg state we implement coherent population trapping (CPT) with respect to the target atoms which were on the ground state initially \cite{2007Fundamentals}. The Rydberg excitation of target atoms will return a significant change to the EIT absorption signal of the  surrounding probe atoms. The transmitted probe light on the surrounding probe atom serves for detection of target positions under the condition of off-resonant EIT.

In a practical experiment, 
given the ground atoms one has to prepare all target atoms onto their Rydberg states through repeated excitations in the CPT configuration. Numerical description for the multi-target-atom Rydberg preparation has been described in Appendix A. Here we simply consider the case of single target atom at a stationary position $r_a$ which was pre-prepared on the Rydberg state $|r_a\rangle$ before detection.
The presence of such an excited target atom can modify $\delta(r)$ of the probe atoms, as defined in Eq.(\ref{drr}), returning an enhanced transmission signal for the probe light. It happens when $|r-r_a|=R_0$ is met.
In the numerical simulation we assume $\{\rho_{a,rr}\}\equiv 1$ denoting its pre-excitation. Denotations $\{\rho_{a,rr}\}\equiv 1$ and 0 means that the target atom is on $|r_a\rangle$ or not.
Other parameters are as follows:
a quasi-2D atomic ensemble with ultracold $^{87}Rb$ atoms at $T_0=10$ $\mu$K, serves as the probe atoms. The peak probe atomic density is $n_0=5\times10^{11}$ cm$^{-3}$ and radially follows a Gaussian density distribution with a half-width $\sigma_r=0.7$ mm \cite{PhysRevLett.107.213601}. The unique
target atom placed in the center is also rubidium atom. The specific experimental energy levels are $|g_b\rangle=|5S_{1/2}, F=2, m_F=2\rangle$, $|e_b\rangle=|5P_{3/2}, F=3, m_F=3\rangle$, and \textcolor{black}{$|r_b\rangle=|80S_{1/2}\rangle$} for the probe atom; and \textcolor{black}{$|r_a\rangle=|85S_{1/2}\rangle$} for the target atom. \textcolor{black}{The {\it vdWs} coefficient is $C_6^{ab}/2\pi=204.8$ GHz$\cdot\mu m^3$ (calculated by \cite{SIBALIC2017319})}. Coefficients $\gamma_2$, $\gamma_3$ are contributed by the decay rates of $|e_b\rangle$, $|r_b\rangle$ and additional dephasing rates. 
The decay rates are estimated to be \textcolor{black}{$\Gamma_{eg}/2\pi=6.0$ MHz and $\Gamma_{re}/2\pi=0.28$ kHz} corresponding to energy levels \cite{PhysRevA.79.052504}. The dephasing effect $\gamma_{e,r}$ comes from nonradiative collisions, Doppler shifts, inhomogeneous trapping potential, and the excitation laser linewidth \cite{PhysRevA.87.053414}. \textcolor{black}{The total damping rates are chosen to be $\gamma_2/2\pi=6.1$ MHz and $\gamma_3/2\pi=10$ kHz}. \textcolor{black}{To ignore the interaction between probe atoms, we require the maximal density on state $|r_b\rangle$ below 1.0 within the blockade radius of $R_c^{\prime}$, {\it i.e.} $\frac{\Omega_{pb}^2}{\Omega_{cb}^2}nL_z\pi R_c^{\prime2}<1.0$ \cite{PhysRevLett.108.013002}. Here $R_c^{\prime}=5.7$ $\mu$m is for the probe-probe blockade}. For realizing an EIT excitation the Rabi frequencies of the probe atom are chosen to be $\Omega_{pb}/2\pi=0.15$ MHz and $\Omega_{cb}/2\pi=15$ MHz.

With the above experimentally-accessible parameters, we transform the absorption into the frame of probe transmission and obtain \cite{RevModPhys.77.633}
 \begin{eqnarray}
T(x,y)=|\exp\{ik_pL_z\chi(x,y)/2\}|
\label{trans}
\end{eqnarray}
in order to show the transmission spectra where $k_p=|\vec{k}_p|$ is the probe wavevector and $L_z=10$ $\mu m$ is the $z$-axis thickness of the medium. For realizing a quasi-2D system we use $L_z\ll L_{x,y} $. In plotting transmission $T$ we reduce it into the form of
\begin{equation}
  T=\exp[i\eta e^{-\frac{r^2}{2\sigma_r^2}}\rho_{b,eg}(x,y)]
\end{equation}
with a dimensionless pre-coefficient \textcolor{black}{ $\eta=k_pL_z\mu_{ge}n_0/\epsilon_0E_p\approx29.19$ and $\mu_{ge}=2.5\times10^{-29}$ C$\cdot$m}.

As displayed in Fig.\ref{fig:singleatom}\textcolor{black}{(a1-a2)} the transmission signal of single target atom is represented by one-cycle measurement within an amplified area of $(15\times 15)$ $\mu$m$^2$. Each pixel is set to be a \textcolor{black}{$(0.25$ $\mu m)^2$}-region to fit the width of transmission ring. For $\Delta_{cb}=0$ the location of the unique target atom denoted by white cross, can be resolved in a broad  disk area with its radius \textcolor{black}{$R_c\approx4.7$ $\mu$m}, which is equivalent to a typical blockade range. However if a nonzero \textcolor{black}{$\Delta_{cb}/2\pi=60$ MHz} is applied, as shown in (a2), one can easily envisage higher resolution given by an improved transmission signal with a spatial extent $\Delta x$ smaller than $R_c$ by one order of magnitude, by which a nondestructive detection for the central target atom is achievable. That fact is made possible by the exact compensation of the probe-target interaction $U_{ab}$ for the off-resonant detuning $\Delta_{cb}$. \textcolor{black}{In the calculation we have also taken account into the probe-probe interaction $U_{bb}$ as shown in Fig.\ref{fig:ubb}(c), and it
causes a bit poor brightness as compared to the result shown in Fig.\ref{fig:singleatom}(a2). Because Fig.\ref{fig:ubb}(c) is obtained by averaging over 1000 stationary one-cycle measurements with random probe-atom excitations.} If the probe-atom excitation is blocked by a big $U_{bb}$ the target atom cannot be detected at this position, which makes the ring brightness lowered.

Therefore, our numerical results(Fig. \ref{fig:singleatom}) based on stationary one-cycle measurement of a pre-excited target atom, show that both contrast and spatial resolution in IETI approach are greatly improved if compared to the result of IEI with $\Delta_{cb}=0$.
Surely, a higher spatial resolution is principally satiable by a growing $\Delta_{cb}$, \textcolor{black}{however the optical diffraction limit sets an obstacle to it which is uneasy to overcome without auxiliary techniques \cite{kumar_duan_hegde_koh_wei_yang_2012,wang_guo_li_luk'yanchuk_khan_liu_chen_hong_2011,nature7807}.}
So in order to obtain an experimentally-accessible image we will adopt the optimal parameters in \textcolor{black}{(a2)} for studying the many-atom case.
In Figure \ref{fig:singleatom}(b) we verify that, the ratio $\Delta x/R_c$ characterizing the improved strength of the spatial resolution, can be made even smaller via an adjustment of $\Delta_{cb}$. \textcolor{black}{And at the same time the annulus radius $R_0$ in IETI can also be shrunk into much smaller than the blockade radius $R_c$ as long as $\Delta_{cb}$ is appropriately chosen.}

\subsection{Fast response in real-time imaging}
\label{fast}

Nondestructive detection requires a repeated measurement for Rydberg targets; however once the control or probe light switches with time, 
the detected system needs a finite response time for reaching a new equilibrium. This response speed of Rydberg EIT could qualify the property of the real-time imaging. So far a fast response property of Rydberg EIT has been verified to be facilitated by the presence of strong Rydberg-Rydberg interactions \cite{PhysRevA.97.043821,PhysRevA.101.023806}. We numerically study the real-time absorption behavior of the probe atoms at a determined distance $R_0$ which could reflect the least time required for one-cycle measurement.
The initial probe atoms are assumed to undergo a sudden switch-on of the coupling light $\Omega_{cb}(t)$, consequently attaining a stationary state during the measurement time. But once the coupling laser is switched off after stable measurement $T_{meas}$, the system tends to recover accompanied by a different recovering time. We can roughly estimate the least time required for a complete one-cycle measurement which should contain three processes of response, recover and stationary measurement.
A faster response time could improve the imaging quality, accelerating real-time detection in experiment.

\begin{figure}
\centering
\includegraphics[width=3.7in,height=2.3in]{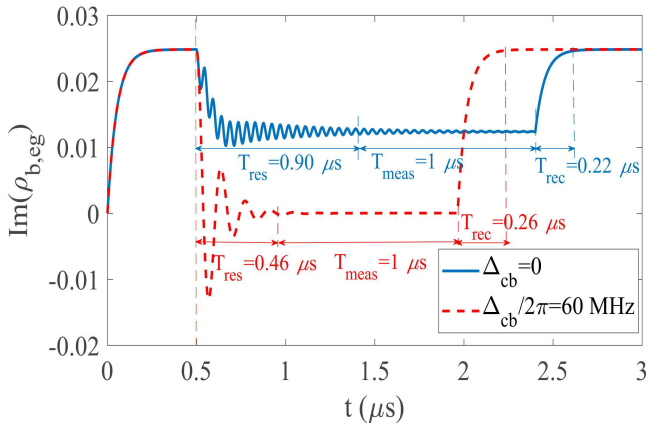}
\caption{Transient behavior of the probe absorption $Im(\rho_{b,eg}(t))$ as a function of time $t$ undergoing a sudden switch-on of the coupling laser at $t=0.5$ $\mu$s and
a sudden-off of it after a $1$-$\mu$s stationary measurement {\it i.e.} $T_{meas}=1.0$ $\mu$s. By using $\Delta_{cb}/2\pi=(0,60)$ MHz, the total imaging time which contains response time $T_{res}$, stable measurement $T_{meas}$, recover time $T_{rec}$, are separately denoted.}
\label{fig:time}
\end{figure}

Numerical results are obtained based on solving the time-dependent dynamics of $Im(\rho_{b,eg})$ from the single-atom master equation (2) without the probe-probe interactions. We switch on the probe field $\Omega_{pb}$ at $t=0$ and preserve it unchanged, leading to \textcolor{black}{$Im(\rho_{b,eg})\to \Omega_{pb}/\gamma_2\approx 0.025$}. At $t=0.5$ $\mu$s we open the coupling field $\Omega_{cb}$ to study the transient response of the probe absorption, which gives rise to a new stationary solution $Im(\rho_{b,eg})\to 0$ at the EIT window $R=R_0$ if $\Delta_{cb}\neq0$. 
As for $\Delta_{cb}=0$ we show the dynamics at $R=R_c$. After a $1.0$-$\mu$s measurement, the control field is switched off again. Figure \ref{fig:time} comparably shows the transient behavior at on-resonance or at far-off resonance. The probe absorption $Im(\rho_{b,eg}(t))$ tends to be stationary onto new status with different speeds after the first switch-on of the coupling $\Omega_{cb}(t)$ at $t=0.5$ $\mu$s. It is clear that before the switch-on of $\Omega_{cb}(t)$, $Im(\rho_{b,eg}(t))$ is same for all cases due to the decoupled state $|r_b\rangle$, \textcolor{black}{reducing to a two-level scheme}. However once $\Omega_{cb}(t)$ is present the frequency shift of $|r_b\rangle$ will strongly impact the probe absorption. For $\Delta_{cb}=0$ this shift solely caused by the strong probe-target interaction $U_{ab}$, will lead to a long response time with strong oscillations towards the steady state, typically around $T_{res}=0.90$ $\mu$s. Fortunately when $U_{ab}$ is exactly overcome by a finite $\Delta_{cb}$ as applied in our scheme, an effective resonant excitation with $\Delta_{cb}^{\prime}=0$ for the probe atoms could favor a faster response time $T_{res}=0.46$ $\mu$s to be stationary. A numerical criterion for stationary state in the calculation is estimated by an average fluctuation of absorption within a time period of $0.1$ $\mu$s that meets the condition of $|\delta Im(\rho_{b,eg})|<10^{-3}$. Note that the response behavior is totally regardless of the exact $\Delta_{cb}$ values because the perfect compensation of $\Delta_{cb}$ by a suitable probe-target interaction $U_{ab}$ at a distance $R_0$, will cause a same resonant two-photon excitation for the probe atoms.
So different $\Delta_{cb}$ values give rise to exactly same transient behaviors. Due to the fast response time by an off-resonance spatial EIT we can safely assume that the atoms are nearly stationary under the cryogenic environment.

After the fast EIT response we set a same time period for carrying out the stable measurement {\it i.e.} $1.0$ $\mu$s refers to the duration of one-cycle imaging. Then the coupling field $\Omega_{cb}(t)$ is turned off again, arising a similar recovering time back to the original status. Therefore an one-cycle measurement requires totally $T_{res}+T_{meas}+T_{rec}\approx \textcolor{black}{1.72}$ $\mu$s (this value becomes 2.12 $\mu$s in the IEI case) which could be also regarded as the temporary resolution of our scheme. \textcolor{black}{Repeated measurements via a time-dependent coupling field $\Omega_{cb}(t)$ can be performed on the system where a second $1$-$\mu$s measurement would have a perfect agreement with the first measurement since they start from same initial status.
Compared to the on-resonant IEI scheme our method can save the time for one-cycle measurement due to its faster response which enables more repetitive measurements within the duration of Rydberg-state lifetime.}
To our knowledge in a real-time imaging process such measurement with fast response can be performed by monitoring the stationary images using a suitable detector, {\it e.g.} EMCCD \cite{PhysRevLett.124.053401}. This detector has shown its preeminent ability in the weak-field measurement because of its ultralow noise, high resolution, high-quantum efficiency, and the robustness to overexposure \cite{1210766}.



\section{Many-atom imaging simulation}
\label{mainresults}
\begin{figure*}
\centering
\includegraphics[width=5.0in]{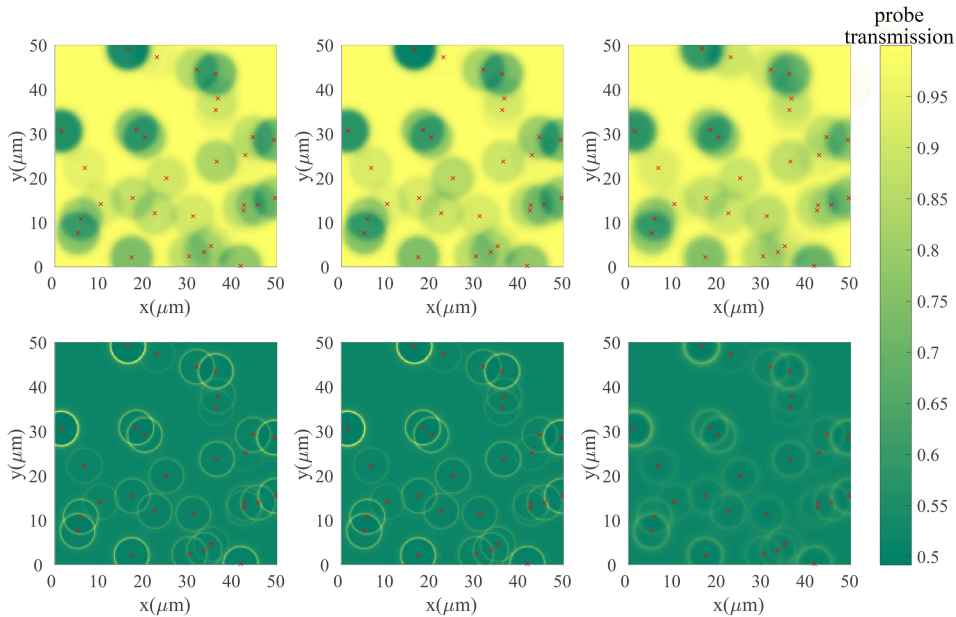}
\caption{(color online) Simulated many-atom transmission images with 30 randomly-distributed target atoms within the area of $(50\times 50)$ $(\mu m)^2$. \textcolor{black}{The red crosses indicate the initial position of target atoms. From top to bottom by increasing the detuning $\Delta_{cb}/2\pi=(0,60)$ MHz}, the transmission images for many atoms become more and more distinguishable with higher spatial resolution, accounting for the dual realization of $R_0<R_c$ and $\Delta x\ll R_c$ in IETI. \textcolor{black}{From left to right the thermal motion of target atoms are involved under the temperature $T_0$ of $0$, $10$ $\mu$K and 1 mK}. \textcolor{black}{The target atoms are pre-excited before stationary detection. Sufficient iterations $n_{max}$(=1000) are used for preserving a high Rydberg probability of state $|r_a\rangle$ which ensures all targets detectable.} Here \textcolor{black}{$(\Omega_{ca},\Omega_{pa})/2\pi=(1,10)$ MHz} for the Rabi frequencies of target excitation and the pre-coefficient  \textcolor{black}{$C_6^{aa}/2\pi=-1347$ GHz$\cdot \mu m^6$} used for calculating the {\it vdWs}-type interaction between two target atoms which is \textcolor{black}{$U_{aa}=C_6^{aa}/r_{aa}^6$}. }
\label{fig:global}
\end{figure*}

For the case with more target atoms, we have to perform a many-body quantum simulation following the stationary probability $\rho_{b,eg}(r)$ of each probe atom. Detailed simulation procedures are described in the Appendix A. In Fig.\ref{fig:global} we present the transmission images of probe atoms \textcolor{black}{averaging over sufficient one-cycle measurements} for resolving the critical position of target atom $i$. Note that all probe-target $U_{ab}$ and target-target $U_{aa}$ interactions play roles in the calculation. The former produced by the energy shift of state $|r_b\rangle$ can be overcome via an off-resonance detuning $\Delta_{cb}$, giving rise to a precise positioning of target atoms, while the latter decides the Rydberg-state probability of them. \textcolor{black}{The CPT technique with Rabi frequencies $\Omega_{pa}\gg\Omega_{ca}$ ensures a higher excitation probability on average}.
Once they are kept sustaining on state $|r_a\rangle$ it leads to  $\{\rho^i_{a,rr}\}=1$, in analogy to the single target atom case where the atomic position could be precisely resolved. During one-cycle measurement, after preparation we preform a $1.0$-$\mu$s stationary measurement. Due to the low temperature as well as the fast response, we safely assume that all target and probe atoms have been stationary. Here we simply use 1.0 $\mu$s as the measurement time.

 However due to the finite temperature $T_0$ which leads to the thermal motion of Rydberg atoms in a real implementation, it may make the transmission ring very blurred. Taking account into the atomic motion we consider all randomly-embedded target atoms move stochastically whose velocities satisfy a two-dimensional Maxwell-Boltzmann distribution
\begin{equation}
  f(v_x,v_y)=\frac{m}{2\pi k T_0}\exp[-\frac{m(v_x^2+v_y^2)}{2kT_0}]dv_xdv_y 
\end{equation}
with $k$ the Boltzmann constant, $m$ the atomic mass and $v_{x(y)}$ the velocity along $\hat{x}(\hat{y})$. Note that we have realized a Doppler-free measurement for probe atoms by letting the probe and coupling fields near-perpendicular to the probe atomic motion. To numerically estimate this effect, the position $\vec{r}_i$ of every target atom $i$ is determined by 
\begin{equation}
    \vec{r}_i \to \vec{r}_i +(v_{x}\vec{x} + v_y\vec{y})t
    \label{ri}
\end{equation}
where $v_{x}$ and $v_{y}$ are obtained stochastically from the velocity distribution function $f(v_x,v_y)$ characterizing the average thermal speed of target atom $i$ during $T_{meas}= 1.0$ $\mu$s. The variation of $|r_i|$ can modify the probe-target interaction so as to change the radius $R_0$ of the transmission ring. Our results with 30 target atoms are based on an average over 1000
one-cycle measurements, taking into account all imperfections from atomic movement. Each measurement contains a repetitive Rydberg excitation of target atoms and a stationary measurement for the probe transmission through EIT spectra. \textcolor{black}{Given moving target atoms, the simulated images are obtained by following an integral averaging over the measurement time, as
\begin{equation}
    T(x,y) = \frac{1}{T_{meas}}\int_0^{T_{meas}}T(x,y,t)dt
\end{equation}
where the stationary target position $\vec{r}_i$ in probe transmission $T(x,y,t)$ is replaced by $\vec{r}_i+\vec{v}t$.
It is intuitive that our detection works well in a lower temperature $T_0$ where the movement of target atoms can be negligible. While a higher $T_0$ would make the transmission ring drifting and blurred.}
In addition to the thermal motion of atoms, heating effect originated from the fluctuation of the laser intensity may also lead to the imperfection of imaging \cite{PhysRevA.61.045801}. Nevertheless in our scheme a weak probe laser at off-resonance gives rise to a poor excitation probability which makes the heating effect less important.

Figure \ref{fig:global} globally illustrates the transmission images of random 30 atoms distributed over a wider regime of $(50\times50)$ $\mu m^2$ with \textcolor{black}{$(200\times200)$} pixel points. Each pixel has an area of $(0.25$ $\mu m)^2$. In the way of traditional IEI [the first row], every target atom can be resolved by detecting the lower-transmission window of the probe EIT, yet suffering from a big reduction of spatial resolution if $N_A$(the number of target atoms) is large. Because the image spots decided by the blockade radius $R_c$ significantly overlap in the space as $N_A$ becomes larger which leads to a poor resolution. For example, in the first row of Fig.\ref{fig:global} each target atom is imaged by a broad disk [see Fig.\ref{fig:singleatom}(a1) for single-atom imaging] that the whole image quality becomes very poor due to the overlap among different
low-transmission disks. Here different temperatures give rise to less-distinguishable changes because the imaging disk is so broad that a small movement of the central target atom has no visible impact on the imaging pattern. \textcolor{black}{Even at $T_0=1$ mK the most probable speed of target atom is $v_{mps}\approx 0.4$ $\mu$m/$\mu$s leading to a drifting distance of 0.4 $\mu$m which is much smaller than the disk radius $R_c$.}

In our new protocol IETI, by tuning $\Delta_{cb}$ to be a nonzero value the quality for atomic imaging obtains a great improvement as shown in the second row of Fig.\ref{fig:global}. Here $\Delta_{cb}/2\pi=60$ MHz is used. 
\textcolor{black}{That fact is mainly caused by higher-probe transmission in IETI accompanied by a flexible target-probe spacing $R_0$ that depends on $\Delta_{cb}$, which leads to a narrow transmission ring for determining the position of all targets. Surely the virtue of IETI also lies in that both the ring radius $R_0$ and its fluctuating extent $\Delta x$ can be freely tuned by the detuning $\Delta_{cb}$. Therefore compared to IEI scheme in which $\Delta_{cb}=0$, the presence of a narrow and sharp transmission ring can achieve a higher-resolution imaging of target atomic positions in IETI scheme}. 
On the other hand, when the temperature $T_0$ increases to $10$ $\mu$K in which $v_{mps}\approx 0.04$ $\mu m/\mu s$, no visible changes of the image can be found. Because during measurement, the movement of each target atom $i$ is only 0.04 $\mu m$ which is smaller than the ring thickness $\Delta x$ by one order of magnitude.
Only if $T_0$ is increased to be 1 mK which is higher than a typical experimental temperature, a moving target atom will fluctuate the probe-target interaction $U_{ab}$, making the image slightly faint because $v_{mps}T_{meas}\approx \Delta x$.
\textcolor{black}{A rough estimation for a maximal temperature that our scheme works is about 1 mK because if the drifting distance of transmission ring is larger than the ring width {\it i.e.} $v_{mps}T_{meas}>\Delta x$ it is impossible to determine its central position with a high precision.}
Also, we see the image contrast on the ring suffers from a clear reduction.
The reason comes from an average over 1000 one-cycle measurements in which the slim ring is sensitive to the position of target atoms {\it i.e.} the slim ring would be drifted in every measurement. As a consequence the final detected image becomes lower-contrast and blurred as the temperature increases.

\section{Feasibility discussion and Conclusion}

 Now we discuss the scheme feasibility by assessing the signal-to-noise ratio($SNR$) during the IETI measurement. The noise in the measurement typically comes from the shot noise of photons and of atoms respectively. They are described by a variance of the transmission signal $\Delta T$ \cite{PhysRevLett.108.013002}
 \begin{equation}
     \text{var}(\Delta T)\approx \frac{2+\langle T\rangle+\langle T\rangle^2}{\langle N_r\rangle} + \frac{\sigma_0^2nL_z}{a}\text{Im}(\chi)^2\langle T\rangle^2
     \label{var}
 \end{equation}
where the first term attributes to the Poisson distributed photon-shot noise and the second term is from probe atomic density fluctuations. $\Delta T$ is the relative transmission rate with respect to that from a reference image generated in the case of no target atoms. $\sigma_0$ is the probe absorption cross-section. To ensure that the density of the Rydberg probe atoms is kept low and the EIT condition $\Omega_{pb}\ll\Omega_{cb}$ is satisfied, we assume $\Omega_{pb}^2=0.1\Omega_{cb}^2n_0L_zA$. In the limit of the strongest transmission $\langle T\rangle\approx1.0$ we assume the mean photon number at each pixel is $\langle N_r\rangle \approx 0.1at_{exp}\Omega_{cb}^2/\sigma_0n_0L_z\Gamma_{eg}A$ with $A=2\pi R_0\Delta x$ for the ring area, $n_0$ the peak density, $a$ the area of each pixel and $t_{exp}$ the exposure time determined by the stable measurement time $T_{meas}$. Hence Eq.(\ref{var}) can be explicitly rewritten as
\begin{eqnarray}
\text{var}(\Delta T) &=&
\frac{\sigma_0\Gamma_{eg}n_0L_zA}{10a\Omega_{cb}^2t_{exp}} \times \nonumber\\
&(4&+\frac{\Omega_{cb}^2t_{exp}\sigma_0\text{Im}(\chi)^2}{A\Gamma_{eg}}e^{-2\sigma_0n_0L_z\text{Im}(\chi)})
\label{vart}
\end{eqnarray}
with Im$(\chi)=(1+0.2\Omega_{cb}^2/\Gamma_{eg}AnL_z)^{-1}$ characterizing the probe absorption. One can define $\text{SNR} = \Delta T/\sqrt{\text{var}(\Delta T)}$, which has to be larger than 1 for detecting the target atoms in the imaging. 

\begin{figure}
\centering
\includegraphics[width=3.6in,height=1.6in]{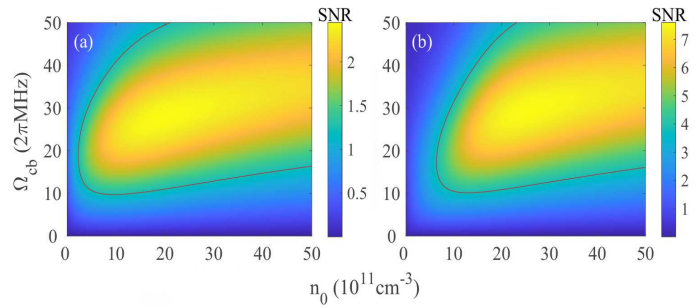}
\caption{Estimated $SNR$ with the parameters $(n_0,\Omega_{cb})$ when the exposure time is (a) $t_{exp} = 1$ $\mu$s and (b) $t_{exp} = 10$ $\mu$s. The solid line in each plot depicts the contour of $SNR=1$. Other systematic parameters are \textcolor{black}{$a=0.16$ $(\mu m)^2$, $\sigma_0=0.145$ $(\mu m)^{2}$, $\Gamma_{eg}/2\pi = 6.1$ MHz, $A=10$ $(\mu m)^2$, $L_z=10$ $\mu$m and $T_0=0$ K.}}
\label{fig:snr}
\end{figure}

\textcolor{black}{In Figure \ref{fig:snr}(a-b) the numerical estimation for SNR distribution is represented when both
the peak atomic density $n_0$ and the coupling strength $\Omega_{cb}$ are adjusted for different exposure times.} \textcolor{black}{Note that the SNR is relatively smaller as compared to the earlier IEI schemes \cite{2016Interaction}, since we require smaller pixels to image the slim transmission ring, which may add a larger shot noise due to $\text{var}(\Delta T)\propto a^{-1}$. Here we choose each pixel containing the width of ring and hence the pixel area is $a=(\Delta x)^2\approx(0.4$ $\mu m)^2$, following the parameters in Fig.\ref{fig:singleatom}(a2).} By comparing the simulation results of Fig.\ref{fig:snr}(a-b) where the atomic movement is ignored, we observe that for $t_{exp}=1$ $\mu$s, the optimal SNR sustains above 1.0 at appropriate $n_0$ or $\Omega_{cb}$ values. And a longer-time measurement $t_{exp}=10$ $\mu$s will arise a larger SNR because of the continuous excitation of probe atoms. 
Extra imperfections such as atomic thermal motion brought by a longer exposure time will leave for consideration in the future.
A rough estimation of the mean photon number based on each pixel arises \textcolor{black}{$\langle N_r\rangle\approx0.5$} for $n_0=5\times10^{11}$ cm$^{-3}$ and $t_{exp}=1$ $\mu$s.
\textcolor{black}{Another obstacle arises with the increase of probe atomic density $n_0$, which is the avalanche effect. This effect is possible to excite more nearby probe atoms to the Rydberg state, destroying an accurate positioning of every target atom. Luckily it is difficult to observe the avalanche excitation since the maximal probe excitation rate $\rho_{b,rr}^{\max}\propto(\Omega_{pb}/\Omega_{cb})^2=10^{-4}$ is very low even at resonance, which can not reach the critical point as reported in Refs. \cite{PhysRevX.10.021023,nature577}.}

In conclusion, we propose an improved IETI technique for the nondestructive determination of atom positions with both high spatial resolution and fast response time. This increased resolution compared to previous IEI method mainly attributes to the use of an off-resonant spatial EIT excitation with respect to the background probe atoms. Once the probe detuning of Rydberg levels suitably overcomes the strong probe-target interaction, it induces \textcolor{black}{an extremely narrow absorption dip} at a critical probe-target distance. \textcolor{black}{One can precisely resolve the location of the target atoms by using this narrow window, attaining a spatial fluctuation about 0.4 $\mu$m constrained by the optical diffraction limit}. Furthermore, a higher resolution is expected in the IETI approach via the adjustment of detuning solely. 
The IETI approach, not only preserves the merits of traditional IEI method that is a nondestructive and state-sensitive technique, but also promises a challenge of superresolution Rydberg imaging ascribed to the new control knob from an off-resonance EIT excitation to the background probe atoms. The response time for single measurement also obtains a great improvement in IETI which is suitable for a present practical performance. All the parameters we use to optimize the simulated images closely meet with the current experimental conditions, deserving for an experimental exploration with cold Rydberg atoms in a practical ensemble.

To exceed the diffraction limit determined by the imaging laser field, existing techniques often utilize point-to-point image reconstruction with multiple measurements per point by scanning or labeling \cite{gustafsson_2000,Rust2006Sub}. For example, stochastic optical reconstruction microscopy (STORM), is eligible to be applied here to overcome this limit. The principle of STORM is utilizing a few photons for each shot of imaging but consecutively sufficient photons to enable precise localization \cite{Rust2006Sub}. Eqs.(\ref{rhoeg}) and (\ref{chir}) imply that the EIT spatial imaging is independent of the intensity of the probe laser and therefore the IETI technique works with low-light-level probe imaging beam. With photon number of the probe laser reduced to a few photons, we expect that the best spatial resolution reaches tens of nanometers. 
Moreover, machine learning(ML) can accomplish the task of classification and identification of complex image patterns, so it can robustly improve the process of image recognition by providing training data based on resembling experimental data, not on idealized theoretical predictions \cite{ghosh_roth_nicholls_wardley_zayats_podolskiy_2021}.
Optical reconstruction techniques based on ML algorithm may initiate another fast and accurate characterization of object structures and provide a promising approach for our scheme to reach superresolution many-atom imaging in the future
\cite{Li:21}.


\bigskip

\acknowledgements

 This work is supported by the National Key Research and Development Program of China under Grants No. 2016YFA0302001 and No.2019ZT08X324; by the National Natural Science Foundation of China under Grants No. 12174106, No.11474094, No.11104076 and No.11654005; by Guangdong Provincial Key Laboratory under Grant No.2019B121203002; by the Science and Technology Commission of Shanghai Municipality under Grant No.18ZR1412800; by the Shanghai Municipal Science and Technology Major Project under Grant No. 2019SHZDZX01; and also additional support from the Shanghai talent program.

\appendix


\section*{Appendix A: Quantum simulation}
\label{manyatoms}


{\it Rydberg preparation of multi-target atoms.}
The transmission imaging signal can be numerically simulated via a semiclassic Monte Carlo approach \cite{PhysRevA.88.043431}. \textcolor{black}{The target energy levels as presented in Fig.\ref{fig:model}(b), adopt a two-photon excitation with $|g_a\rangle=|5S_{1/2}\rangle$, $|e_a\rangle=|5P_{3/2}\rangle$ and $|r_a\rangle=|85S_{1/2}\rangle$. The optical couplings among them are $\Omega_{ca}/2\pi=1$ MHz, and $\Omega_{pa}/2\pi=10$ MHz.} Given the initial status with all target atoms in the ground states $|g_a\rangle$ we assume an initial one-dimensional array for denoting their status
\begin{equation}
  \{\{\rho^i_{a,rr}\}\}_{n=0}=\{0,0,0,...,0,0,...\} 
\end{equation}
where the superscript {\it i} represents the {\it i}th embedded target atom, and denotation $\{\rho^i_{a,rr}\}=1$ or $0$ means that the target atom is excited to $|r_a\rangle$ or not. The stationary Rydberg probability $\rho^i_{a,rr}$ of the $i$th atom is given by \cite{PhysRevA.87.023401}
\begin{equation}
    \rho_{a,rr}^i=\frac{\left|\Omega_{pa}\right|^2(\left|\Omega_{pa}\right|^2+\left|\Omega_{ca}\right|^2)}{(\left|\Omega_{pa}\right|^2+\left|\Omega_{ca}\right|^2)^2+(\gamma^2_{ea}+2\left|\Omega_{pa}\right|^2)\delta_i^{(n)}}
\end{equation}
where $\gamma_{ea}/2\pi=6.1$ MHz. $\rho_{a,rr}^i$ varies with the two-photon detuning $\delta_i^{(n)}$ and for $\delta_i^{(n)}=0$ it attains a maximal value.

Initially $n=0$ we assume a resonant two-photon detuning  $\delta_i^{(0)}=\Delta_{pa}+\Delta_{ca}=0$, giving rise to a maximal probability $\rho_{a,rr}^i$. No target-target interaction presents initially.
Next one generates a random number $s_i$ between $(0,1)$ for each target atom $i$. If $s_i\leq\rho^i_{a,rr}$ we set $\{\rho^i_{a,rr}\}=1$(excited), otherwise, $\{\rho^i_{a,rr}\}=0$(not excited). That will arise a new array, {\it for example} $\{\{\rho_{a,rr}^i\}\}_{n}=\{0,1,0,0,...,0,1,0,1,0,...\}$ with $n=1,2,...n_{max}$ representing the iterations. $\{\{\rho_{a,rr}^i\}\}_{n}$ denotes the updated status of all target atoms.

For the $i$th target atom, any other target atom $j$ in the Rydberg state $|r_{a}\rangle$, {\it i.e.} $\{\rho^j_{a,rr}\}=1$, will induce a target-target level shift, translating into the $i$th-atom two-photon detuning $\delta_i$, given by
\begin{equation}
\delta_i^{(n+1)}=\delta_0+\sum_{j\ne i}^{N_A}\{\rho_{a,rr}^{i}\}_n\frac{C_6^{aa}}{|r_j-r_i|^6},
\end{equation}
For the initial step, $n=0$ and  $\delta_{i}^{(0)}=\delta_0$. $N_A$ is the number of target atoms and $\frac{C_6^{aa}}{|r_j-r_i|^6}$ stands for the intraspecies $vdWs$ interactions between $i-j$ target atoms that possess Rydberg $S$-state via a two-photon excitation with $C_6^{aa}/2\pi=-1347.2$ GHz \cite{SIBALIC2017319}.

Finally a large number of target atoms would be prepared on $|r_a\rangle$, giving to the final status of all target atoms, {\it e.g.}
\begin{equation}
  \{\{\rho^i_{a,rr}\}\}_{n_{max}}=\{1,1,1,...,0,1,...\} 
\end{equation}
under sufficient iterations. The maximum $n_{max}$ depends on the excitation parameters used for target atoms. In particular, when two or more target atoms are initially closely placed within the target blockade radius it is difficult to simultaneously excite them owing to the blockade effect. However the probability for that is tiny. 
Therefore, by using sufficient iterations for Rydberg preparation the average excitation probability of $i$th atom finally expressed as
\begin{equation}
 \bar{\rho}_{i}=\frac{1}{n_{max}}\sum_{n=1}^{n_{max}}\{\rho_{a,rr}^i\}_n   
\end{equation}
is very close to 1.0.
Note that if any target atom remains unexcited after $n$ iterations it can not be detected in the imaging process due to the absence of the probe-target interaction.
\textcolor{black}{To avoid this, we repeatedly pump the target atoms from the ground state before detection.
We note that during the detection process these pre-excited target atoms can sustain on the Rydberg state and thus a non-destructive measurement using off-resonant Rydberg EIT, can be performed.}

{\it Detection for transmitted probe light.} Given the pre-excitation status of target atoms we can calculate the stationary probe absorption at off-resonance by following Eq.(\ref{rhoeg}), where the effective two-photon detuning $\delta(r)$ reflecting the probe-target interaction is replaced by \textcolor{black}{
\begin{equation}
\delta(r)=\Delta_{cb}-\sum_{i(r_i\ne r)}^{N_A}\{\rho_{a,rr}^i\}\frac{C_6^{ab}}{|r-r_i|^6},
\label{deltan}
\end{equation}}
For any atom $i$ only the surrounding probe atoms with a suitable relative distance $|r-r_i|$ that leads to $\delta(r)\approx0$, can reveal a sharp absorption dip in the off-resonant EIT spectra. Other unsuited probe atoms can not be detected. Based on the modified Eq.(\ref{rhoeg}) as well as Eq.(\ref{drr}) it arrives at the first-order susceptibility $\chi(r)$ whose imaginary part stands for the probe absorption rate. The probe transmission is proportional to $|\exp(i\chi(r))|$ following Eq.(\ref{trans}).
By plotting the probe-atom transmission in the $(x,y)$ space we can capture the information of pre-excited target atoms. Due to the high-quality images of transmission spectra which contains position-sensitive and great resolution advantages, target atoms can be clearly resolved for single-time measurement. Stable exposure time required by a typical detector is about a few $\mu$s so here {\it i.e.} the measurement time $T_{meas}=1.0$ $\mu$s is assumed \cite{Guenter2013Observing}.

Therefore in our simulation, all target-target interactions coming from repeated excitations of target atoms, do perform only in the pre-excitation process. Once two random target atoms are closely placed within the blockade range for target atoms, this imperfect Rydberg preparation would make measurement failed. To make all target atoms detectable, we have to perform repeated excitation runs before measurement and finally obtain an average excitation probability $\bar{\rho}_i$ which is very close to 1.0.
In addition all probe-target interactions are considered as long as the target atoms are excited. The enhanced transmission signal closing to 100$\%$ emerges at the absorption dip due to
an off-resonant EIT effect of probe atoms. Differing from the previous IEI scheme using resonant EIT, the off-resonant EIT condition favors a narrower transmission window which promises for a higher spatial-resolution image.

Note that in Sec.\ref{mainresults}
the probe-probe interaction has been ignored accounting for the poor exciting probability of probe atoms at the off-resonant EIT condition. Even in the vicinity of the transmission ring that the two-photon detuning is overcome by the probe-target interaction, the weak-driving condition as well as the sufficient measurements can ensure the impact of probe excitation negligible. A detailed discussion for the probe-probe interaction is presented in Appendix B.
In the calculation we adopt sufficient iternations $n_{max}$ ensuring the Rydberg excitation of target atoms. So the final  results involving the target-target as well as the probe-target $vdWs$ interactions, can return accurate positions of randomly-distributed target atoms in a two-dimensional space, as displayed in Figs. \ref{fig:singleatom} and \ref{fig:global}.

\section*{Appendix B: Influence of the probe-probe interactions}
\label{ppinter}

{\it Theoretical derivation of the  probe-probe interactions.} For probe atoms they have an off-resonant excitation suffering from a poor excitation probability. However once the atom is placed near the transmission ring with a certain distance $R_0$ to the target atom, due to strong target-probe {\it vdWs} interaction that may overcome the two-photon detuning $\Delta_{cb}$, the probe excitation becomes resonant. In this appendix we focus on the impact of the probe-probe interactions.

\textcolor{black}{For probe atoms, we consider the following soft-core potential \cite{PhysRevLett.105.135301,PhysRevA.97.023619}
\begin{equation}
U_{bb}=\begin{cases} 
-\frac{3^6C_6^{bb}}{R_c^{\prime6}},  & \mbox{if }0<r^{\prime}\leq \frac{1}{3}R_{c}^{\prime} \\
-\frac{C_6^{bb}}{r^{\prime6}}, & \mbox{if }r^{\prime}>\frac{1}{3}R_{c}^{\prime}
\end{cases}.
\label{soft}
\end{equation}
with $R_c^{\prime}=5.7$ $\mu$m the blockade radius of probe atoms and $r^{\prime}$ the probe-probe distance.} The presence of the probe-probe interaction would modify the effective two-photon detuning $\delta(r)$ in Eq.\ref{drr} with respect to the probe atom, which is 
\begin{equation}
    \delta(r) = \Delta_{cb}-\frac{C_6^{ab}}{R^6} - \int_{0}^{\infty}U_{bb}n\rho_{b,rr}(r^{\prime})d^3\boldsymbol{r}^{\prime}
\label{a1}
\end{equation}

Here the probe atomic density $n$ is a function of position due to the Gaussian profile of atomic cloud, with its peak value $n_0=5\times 10^{11}$ cm$^{-3}$ as described in Eq.(\ref{chir}), and $\boldsymbol{r}^{\prime}$ denotes the relative distance between two probe atoms. The probe-probe interaction coefficient $C_6^{bb}/2\pi=-662.4$ GHz$\cdot\mu m^6$. The average Rydberg excitation probability $\rho_{b,rr}$ of probe atom can be analytically solved by \cite{PhysRevA.87.023401}
\begin{equation}
    \rho_{b,rr}(r^{\prime})=\frac{\left|\Omega_{pb}\right|^2(\left|\Omega_{pb}\right|^2+\left|\Omega_{cb}\right|^2)}{(\left|\Omega_{pb}\right|^2+\left|\Omega_{cb}\right|^2)^2+(\gamma^2_{2}+2\left|\Omega_{pb}\right|^2)\delta(r^{\prime})}
\end{equation}
due to its two-photon excitation feature.

We carry out a rough estimation for the maximal probe-probe interaction $U_{bb}$(the third term in Eq.(\ref{a1})). At resonance $\delta(r)\approx 0$ leading to a maximal Rydberg fraction  $\rho_{b,rr}^{\max}\propto(\Omega_{pb}/\Omega_{cb})^2 = 10^{-4}$ by considering $\Omega_{pb}/2\pi=0.15$ MHz and $\Omega_{cb}/2\pi=15$ MHz in the weak probe regime. By replacing the above parameters, the maximal $U_{bb}$ enabled by the nearest-neighbor probe atoms at resonance, is
\begin{equation}
U_{bb}^{\max} = n_0\rho_{b,rr}^{\max}\int_0^{\Delta x}U_{bb}(r^{\prime})d^3\boldsymbol{r}^{\prime}\approx-1.26\text{ MHz}
\label{maxubb}
\end{equation}
Actually at off-resonant case the influence of $U_{bb}$ can be safely ignored due to $|U_{bb}|\ll\Delta_{cb}$. However as closing to the transmission ring the cancellation of $\Delta_{cb}$ by a suitable probe-target interaction $U_{ab}$ may make the $U_{bb}$ important, see Eq.(\ref{a1}). \textcolor{black}{We will study the impact of probe-probe interactions using two numerical methods.}

{\it Stationary images with probe-probe interactions.}
One way to resolve the single-target-atom images with probe-probe interactions may use similar approach as described in Appendix A. Given the initial status of ground probe atoms on $|g_b\rangle$ we assume
\begin{equation}
  \{\{\rho^i_{b,rr}\}\}_{n=0}=\{0,0,0,...,0,0,...\} 
\end{equation}
where the superscript {\it i} represents the probe atoms at the {\it i}th pixel, and here $60\times60$ and $150\times150$ pixels have been respectively simulated in a $(15\times15)$ $\mu m^2$-2D plane. $\{\rho^i_{b,rr}\}=1$ or $0$ means that the probe atom is excited to the Rydberg state $|r_b\rangle$ or not. The stationary Rydberg probability $\rho^i_{b,rr}$ of the $i$th atom at initial step $n=0$ is expressed as
\begin{equation}
\rho^i_{b,rr}\approx\frac{\left|\Omega_{pb}\right|^2(\left|\Omega_{pb}\right|^2+\left|\Omega_{cb}\right|^2)}{(\left|\Omega_{pb}\right|^2+\left|\Omega_{cb}\right|^2)^2+(\gamma^2_2+2\left|\Omega_{pb}\right|^2)\delta_0^2},
\label{sigmaarrb}
\end{equation}
due to a two-photon excitation, where a resonant two-photon detuning $\delta_0=\Delta_{cb}+U_{ab}$ is considered. No probe-probe interaction presents at the initial time.

 \begin{figure}
 \centering
\includegraphics[width=3.5in,height=2.3in]{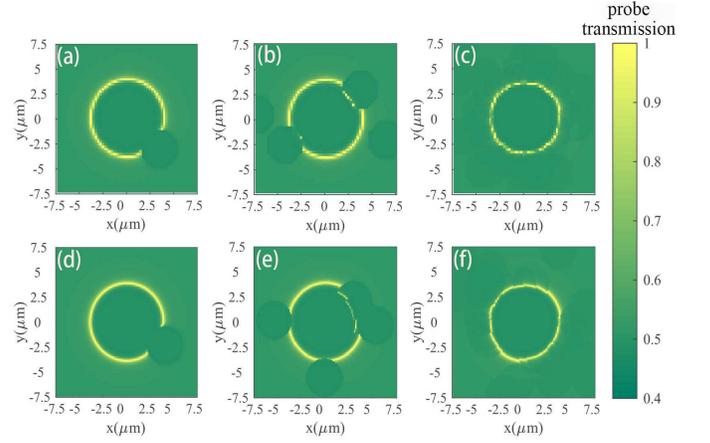}
\caption{Numerical representation of the single-target atom image with probe-probe interactions, covering the range of $(15\times 15) $ $\mu m^2$ for (a-c) 60$\times$60  pixels and (d-f) 150$\times$150 pixels. Areas of each pixel are $(0.25$ $\mu m)^2$ and $(0.1$ $\mu m)^2$ respectively. From left to right different iterations(repeated measurements) are used which are 10, 100, 1000. Other parameters are adopted from Fig.\ref{fig:singleatom}(a2).}
\label{fig:ubb}
\end{figure}

The probability denoting the probe-atom excitation within the {\it i}th pixel, is computed as
\begin{equation}
    P^i_{rr}=nL_zdxdy\rho^i_{b,rr}.
\end{equation}
where $N_b=nL_zdxdy\ll 1$ is the average number of probe atoms in each pixel. $L_z$ is the $z$-axis thickness of atomic  medium, and $dx$ and $dy$ is the $x$-axis and $y$-axis length of the pixel, with $n$ the number density of probe atoms. No more than one probe atom can be excited to the Rydberg state $|r_b\rangle$ within the same pixel due to the blockade effect.

Next, one generates a random number $s_i$ between $(0,1)$ for each pixel $i$. If $s_i\leq$ $P^i_{rr}$ we set $\{\rho^i_{b,rr}\}=1$(excited), otherwise, $\{\rho^i_{b,rr}\}=0$(not excited). That will give rise to a new array of $\{\{\rho_{b,rr}^i\}\}_{n}$ with $n=1,2,...n_{max}$ representing the iterations. $\{\{\rho_{b,rr}^i\}\}_{n}$ denotes the updated status of all probe atoms.
For the probe atom in the $i$th pixel, if any other probe atom in the $j$th pixel is in the Rydberg state $|r_{b}\rangle$, {\it i.e.} $\{\rho^j_{b,rr}\}=1$, it will induce a probe-probe level shift, translating into the $i$th-atom two-photon detuning $\delta_i$ given by
\begin{equation}
\delta_i^{(n+1)}=\delta_0+\sum_{j\ne i}^{N_B}\{\rho_{b,rr}^{i}\}_nU_{i,j}
\end{equation}
with $N_B$ the number of pixels and $U_{i,j}$ the intraspecies $vdWs$ interaction between $i-j$ probe atoms, using the way of soft-core model as Eq.(\ref{soft}).
For the initial step $n=0$, and  $\delta_{i}^{(0)}=\delta_0$.
As a result, by using sufficient iteration the final average probe-probe interactions of each pixel is actually given by
\begin{equation}
    \bar{U}^i_{bb}=\frac{1}{n_{max}}\sum_{n=1}^{n_{max}}\sum_{j\ne i}^{N_B}\{\rho_{b,rr}^{i}\}_nU_{i,j}
\end{equation}
Then the susceptibility and the transmission of each pixel are given by Eq.(\ref{chir}) and Eq.(\ref{trans}).

Figure \ref{fig:ubb} represents the single-atom imaging with the influence of the probe-probe interactions. It is clearly shown that when the iteration is insufficient the image is incomplete see Fig.\ref{fig:ubb}(a-b) and (d-e) due to the probe-probe interaction. Because the accidental excitation of the probe atom in single measurement will lead to the breakdown of the transmission conditions at $R_0$, making transmission ring incomplete. However, after performing sufficient measurements
an average result can also form a complete transmission ring at the expense of a little brightness, see Fig.\ref{fig:ubb}(c).
In addition we observe that a low pixel(precision) could cause the images blurred. However once a sufficient iteration is applied, {\it e.g.} $n_{max}=$1000, accompanied by a precise measurement(high pixel) an expected transmission ring can be re-observed by averaging the influence of the probe-probe interactions. By comparing Fig.\ref{fig:ubb}(c) and Fig.\ref{fig:singleatom}(a2) it confirms that we can approximately ignore $U_{bb}$ in the calculation.

\begin{figure}
\centering
\includegraphics[width=3.5in,height=1.6in]{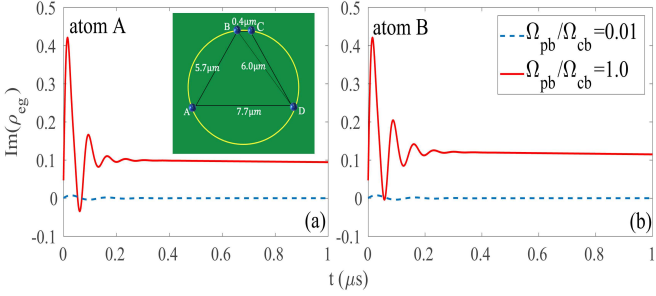}
\caption{\textcolor{black}{(a-b) The absorption $Im(\rho_{eg}(t))$ for two resonantly-excited probe atoms A and B. Results from different Rabi frequencies $\Omega_{pb}/\Omega_{cb}=(0.01,1.0)$, are shown. Here $\Omega_{cb}/2\pi=15$MHz.}}
\label{fig:fewatoms}
\end{figure}

{\it Absorption of a few probe atoms with probe-probe interactions.} \textcolor{black}{To quantitatively discuss the influence of probe-probe interaction we introduce a simple model containing four control atoms at the transmission ring and one target atom in the center, see inset of Fig.\ref{fig:fewatoms}(a). Note that an excited target atom would only lead to a resonant two-photon excitation for probe atoms on the ring. So we can exactly solve the probe-atom evolution by using a few-body master equation excluding the target atom, 
\begin{equation}
    \dot{\rho} = -i[\mathcal{H},\rho] +\mathcal{L}[\rho] 
    \label{mseq}
\end{equation}
where the density matrix $\rho$ becomes a $3^4\times3^4$ matrix. As a result the Hamiltonian composing all atom-field and atom-atom interactions, can be written as
\begin{equation}
\mathcal{H}=-\sum_{i=1}^4(\frac{\Omega_{pb}}{2}\sigma_{eg}^i+\frac{\Omega_{cb}}{2}\sigma_{re}^i+H.c.)+\sum_{i<j}\sigma_{rr}^i\frac{C_6^{bb}}{r_{ij}^6}\sigma_{rr}^j    
\end{equation}
with $\sigma_{\mu\nu}^i\equiv|\mu_i\rangle\langle\nu_i|$ and $r_{ij}$ the probe-probe distance. The Liouville operator $\mathcal{L}=\sum_{i=1}^4(\mathcal{L}_e^i+\mathcal{L}_r^i)$ is expressed as a sum of independent single-atom decay of state $|e_b\rangle$ and dephasing of state $|r_b\rangle$, with $\mathcal{L}_e^i=\frac{\gamma_2}{2}\sum_{i=1}^4(2\sigma_{ge}^i\rho\sigma_{eg}^i-\{\sigma_{ee}^i,\rho\})$ and $\mathcal{L}_r^i=\gamma_3\sum_{i=1}^4((\sigma_{rr}^i-\sigma_{gg}^i)\rho(\sigma_{rr}^i-\sigma_{gg}^i)-\rho)$ \cite{PhysRevA.87.053414}.}

\textcolor{black}{Given the peak atomic density $n_0=5\times10^{11}$ cm$^{-3}$ we assume two atoms B and C are closely placed with a distance of $1/(n_0L_z)^{1/2}\approx0.4$ $\mu$m, in which the probe-probe interaction is dominant. Other atoms are far separated. By solving the master equation (\ref{mseq}) with full probe-probe interactions we could numerically solve the probe absorption $Im(\rho_{eg}(t))$ during the measurement time for atoms A and B. Note that due to the symmetrical structure the behavior of atom D(C) is equal to that of atom A(B). Comparing Fig.\ref{fig:fewatoms}(a) and 8(b) it is explicit that the absorptions of atoms A and B have a perfect agreement if $\Omega_{pb}/\Omega_{cb}=0.01$(blue-dashed) as used in our work. Because the weak excitation of the probe atoms, even at the transmission ring where two-photon resonance is met, will make the impact of probe-probe interaction negligible. Only if $\Omega_{pb}/\Omega_{cb}$ is raised to 1.0(red solid) that leads to a significant excitation of the probe atoms, the steady absorption of atom B will be slightly higher than that of atom A which is caused by different probe-probe strengths felt by atoms A and B. This simulation again confirms that the probe-probe interaction can be safely ignored in our approach.}

\bigskip


%




\end{document}